\documentclass[preprint,10pt]{elsarticle}
\usepackage[utf8]{inputenc}
\usepackage{amsmath}
\usepackage{amsfonts}
\usepackage{amssymb}
\usepackage{graphicx}
\usepackage{subfigure}
\usepackage{ulem}
\usepackage{color}
\usepackage{tikz}
\usetikzlibrary{decorations.pathmorphing,patterns,arrows,shapes.misc}
\usepackage{tikz-3dplot}
\usepackage[nomessages]{fp}
\usepackage[first=0, last=360]{lcg}
\usepackage{ifthen}
\usepackage{csvsimple}
\usepackage{comment}
\usepackage[hidelinks,unicode=true]{hyperref}

\hypersetup{colorlinks=true,
	linkcolor=blue,
	urlcolor=blue,
	citecolor=blue,
	pdfhighlight=/N
}

\makeatletter
\DeclareRobustCommand*\uell{\mathpalette\@uell\relax}
\newcommand*\@uell[2]{
	\setbox0=\hbox{$#1\ell$}
	\setbox1=\hbox{\rotatebox{10}{$#1\ell$}}
	\dimen0=\wd0 \advance\dimen0 by -\wd1 \divide\dimen0 by 2
	\mathord{\lower 0.1ex \hbox{\kern\dimen0\unhbox1\kern\dimen0}}
}

\begin{document}

\begin{frontmatter}

\title{Nonuniversal critical dynamics on planar random lattices with heterogeneous degree distributions}

\author[1]{Sidiney G. Alves}
\ead{sidiney@ufsj.edu.br}
\author[2,3]{Silvio C. Ferreira}
\ead{silviojr@ufv.br}
\author[1]{Marcelo M. de Oliveira}
\ead{mmdeoliveira@ufsj.edu.br}

\affiliation[1]{Departamento de Estatística, Física e Matemática, Universidade Federal de São João Del-Rei,36490-972, Ouro Branco, Brazil}
\affiliation[2]{Departamento de Física, Universidade Federal de Viçosa,36490-972, Viçosa, Brazil}
\affiliation[3]{National Institute of Science and Technology for Complex Systems,22290-180, Rio de Janeiro, Brazil}

\begin{abstract}

The weighted planar stochastic (WPS) lattice introduces a topological disorder that emerges from a multifractal structure. Its dual network has a power-law degree distribution and is embedded in a two-dimensional space, forming a planar network. We modify the original recipe to construct WPS networks with degree distributions interpolating smoothly between the original power-law tail, $P(q)\sim q^{-\alpha}$ with exponent $\alpha\approx 5.6$, and a square lattice. We analyze the role of the disorder in the modified WPS model, considering the critical behavior of the contact process (CP). We report a critical scaling depending on the network degree distribution. The scaling exponents differ from the standard mean-field behavior reported for CP on infinite-dimensional (random) graphs with power-law degree distribution. Furthermore, the disorder present in the WPS lattice model is in agreement with the Luck-Harris criterion for the relevance of disorder in critical dynamics. However, despite the same wandering exponent $\omega=1/2$, the disorder effects observed for the WPS lattice are weaker than those found for uncorrelated disorder.
\end{abstract} 


\end{frontmatter}

\section{Introduction}

Nonequilibrium phase transitions between an active (fluctuating) state and an inactive, absorbing state~\cite{Marro1999,Henkel2008} arise frequently in a broad class of phenomena, such as epidemic spreading~\cite{PastorSatorras2015}, chemical reactions~\cite{deOliveira2004,Ziff1986}, population dynamics \cite{deOliveira2012}, and related fields. Experimental realizations in liquid crystal electroconvection~\cite{Takeuchi2007}, driven suspensions~\cite{Corte2008}, and superconducting vortices~\cite{Okuma2011} also highlight the importance of this kind of transition.

In analogy with equilibrium phase transitions, it is expected that a continuous absorbing phase transition (APT) can be classified into universality classes depending on a few characteristics of the model, namely symmetries, conserved quantities, and dimensionality~\cite{Henkel2008}. Generically, models such as the contact process (CP), which has short range interactions and exhibits a continuous APT, belong to the directed percolation (DP) universality class in the absence of extra symmetries, disorder, or conservation laws~\cite{Janssen1981,Grassberger1982, Henkel2008}. Other robust classes emerge when conservation laws are included~\cite{Henkel2008}.

Of particular interest is how spatially quenched disorder affects the critical behavior of an APT. Quenched disorder was introduced on regular lattices by random deletion of sites~\cite{Noest1988, Moreira1996, Dickman1998, Vojta2006, deOliveira2008, Vojta2009} or by random spatial fluctuations of the control parameter~\cite{Bramson1991, Faria2008, Vojta2005}. In these cases, it was shown that quenched randomness produces rare regions that are locally supercritical, even when the whole system is subcritical~\cite{Vojta2006b}. The lifetime of such active rare regions grows exponentially with the domain size, leading to slow dynamics with nonuniversal exponents for some interval of the control parameter $\lambda_\text{c}(0) < \lambda < \lambda_\text{c}$, where $\lambda_\text{c}(0)$ and $\lambda_\text{c}$ are the critical points of the clean (ordered) and disordered systems, respectively. This extended interval where critical behavior is observed is called Griffiths phase (GP), which can also be observed in the decays towards the stationary state in the supercritical phase~\cite{Vojta2005}, and it was verified in DP models with uncorrelated disorder, irrespective of the disorder strength~\cite{Moreira1996,Dickman1998, deOliveira2008, Vojta2005, Vojta2009}. Another manifestation of relevant disorder is the smearing of a critical phase transition~\cite{Dickison2005,Vojta2006}  where the singular critical behavior given by power-law is replaced by stretched exponential with characteristic scales.

The  Harris’ criterion~\cite{Harris1974} states that uncorrelated quenched disorder is a relevant perturbation if $d\nu_\perp < 2$, where $d$ is the dimensionality and $\nu_\perp$ is the correlation length exponent of the clean model. In the DP class, where $\nu_\perp = 1.096854(4)$, $0.7333(75)$, and $0.584(5)$, for $d = 1, 2$, and 3, respectively \cite{Henkel2008}, the criterion is satisfied for all dimensions $d < 4$, and, consequently, the disorder is predicted to be relevant according to the Harris criterion. Previous works, based on renormalization group methods and numerical simulations, concluded that the activated-disorder behavior belongs to the universality class of the random transverse Ising model~\cite{Igloi2003,Igloi2004}, thus confirming the Harris' criterion predictions. 

A distinct kind of disorder is introduced by random lattices generated by the Voronoi-Delaunay (VD) triangulation~\cite{Okabe2000}, which is a two-dimensional planar graph with a Poissonian distribution of connectivity with an average degree $\overline q = 6$. In this case, the disorder does not change the critical behavior exhibited by the clean CP~\cite{deOliveira2008b,deOliveira2016}, which was, in principle, at odds with the above-mentioned results for the uncorrelated disorder, which led to strong GPs. Barghathi and Vojta~\cite{Barghathi2014} proposed a solution to the puzzle by applying the theory of Luck~\cite{Luck1993}. The Luck-Harris criterion considers that the stability of the critical point is governed by the decay of spatial fluctuations in the local coordination numbers. Averaging the coordination number $q$ over coarse-grained blocks, the corresponding variance $\sigma_q$ decays algebraically with the block size $L_\text{b}$, $\sigma_q\sim L_\text{b}^{-a}$, where $a=d(1-\omega)$ and $\omega$ is the wandering exponent~\cite{Luck1993}. So, the criterion states that quenched disorder is an irrelevant perturbation if the inequality $a\nu_\perp > 1$ holds. In the case of random dilution, the central limit theorem leads to $a = d/2$, and the criterion reduces to the original Harris criterion. Barghathi and Vojta~\cite{Barghathi2014} showed that coordination spatial fluctuations in VD lattices and other disordered planar graphs are not governed by uncorrelated disorder ($a=1$ in $d=2$), but by strong anti-correlations with exponent $a=3/2$ such that VD lattices satisfy the Luck-Harris condition for irrelevance of disorder since $a_\text{c}=1/\nu_\perp = 1.3636<3/2$.
However, Schrauth and collaborators~\cite{Schrauth2018,Schrauth2019,Schrauth2019b}, reported examples that violate the Luck-Harris criterion even using the approach of Barghathi and Vojta~\cite{Barghathi2014}.

A structure presenting topological disorder was proposed by Hassan et al., the weighted planar stochastic (WPS) lattice~\cite{Hassan2010}. In the WPS lattice, the topological disorder emerges from a multifractal structure~\cite{Dayeen2016}, with its coordination number probability distribution following a power law. It was found that the isotropic percolation universality class on the WPS lattice~\cite{Hassan2016} is distinct from the one for all the regular planar lattices~\cite{Hsu1999}. An opinion-dynamic model also exhibits a critical behavior on the WPS different from the one observed in regular lattices~\cite{Liu2018}. Recently, it was shown that the critical behavior of the contact process on the WPS lattice~\cite{Alves2022} is different from that on a regular one, showing that the coordination disorder introduced by this model is a relevant perturbation of the model.

On the other hand, in infinite-dimensional power-law networks, it is known that the critical exponents depend on the degree distribution exponent~\cite{Ferreira2011,Mata2014} consistently: the exponents agree with the heterogeneous mean-field theory, which explicitly takes into account the shape of the degree distribution~\cite{Ferreira2011,Mata2014}. In particular, the critical exponents are equivalent to those of the usual mean-field theory for non-scale-free networks with degree exponent $\alpha>3$, in contrast with the WPS result, where unusual mean-field exponents were found, despite the degree exponent $\alpha\approx 5.6$ out of the scale-free regime.

So, to further investigate the effects of topological planar disorder, we study the critical behavior of the CP in a modified structure based on the WPS lattice. We change the original algorithm of the partitioning process to introduce dual networks with distinct degree distribution exponents $\alpha\ge\alpha_\text{\footnotesize{WPS}}$. We analyze the critical behavior of the CP on these structures, which exhibits critical scaling that depends on the lattice degree exponent. This dependence differs from infinite-dimensional power-law networks and indicates a generalized DP universality class whose exponents depend on both embedding dimension and degree distribution. Finally, the wandering exponent analysis predicts that the introduced disorder is relevant due to nonuniversal critical exponents that differ from the DP universality class. However, the disorder effects are weaker than those usually observed in uncorrelated cases, where extended regions of criticality are observed~\cite{Vojta2006b,Moreira1996,Vojta2009,deOliveira2008}.

The remainder of this paper is organized as follows. In Sec.~\ref{sec:models}, we present the model and methods employed in the work. In Sec.~\ref{sec:result}, we show and discuss our simulation results. Section~\ref{sec:conclu} is to summarize our conclusions.

\section{Models and methods}
\label{sec:models}

\subsection{Modified WPSL}
We start by presenting the standard algorithm for the WPS lattice introduced in Ref.~\cite{Hassan2010}. Consider a square domain. In the first step, it is divided randomly into four rectangular blocks. In the following steps, a block is chosen at random, one at a time, with probability proportional to its area, and partitioned into four new blocks at random. After $n$ iterations, the lattice has $3n + 1$ nodes. Finally, the dual lattice is constructed in the following way: each block is considered a node of the network, and two blocks are connected if they share borders.

The modified WPS algorithm limits the partitioning region around the rectangles' centers within some interval controlled by a parameter $\epsilon\in [0,1]$. At each iteration, a rectangle of sides $(\uell_x,\uell_y)$ is chosen proportionally to its area. The partitioning points $(x,y)$ are chosen at random within the domain $(1-\epsilon)\uell_x/2 < x < (1+\epsilon)\uell_x/2$ and $(1-\epsilon) \uell_y/2 <y <  (1+\epsilon) \uell_y/2$, where $\epsilon$ is the parameter that controls the randomness. Notice that the limits $\epsilon=0$ and $\epsilon=1$ correspond to a square lattice and the original WPS lattice, respectively. By construction, the average area of the blocks is $\langle A\rangle =1/(3n+1)$, which defines a typical average length  $\langle\uell \rangle = 1/\sqrt{3n+1}$.

\begin{figure*}[bh]
	\begin{center}
		\includegraphics[width=0.2\hsize]{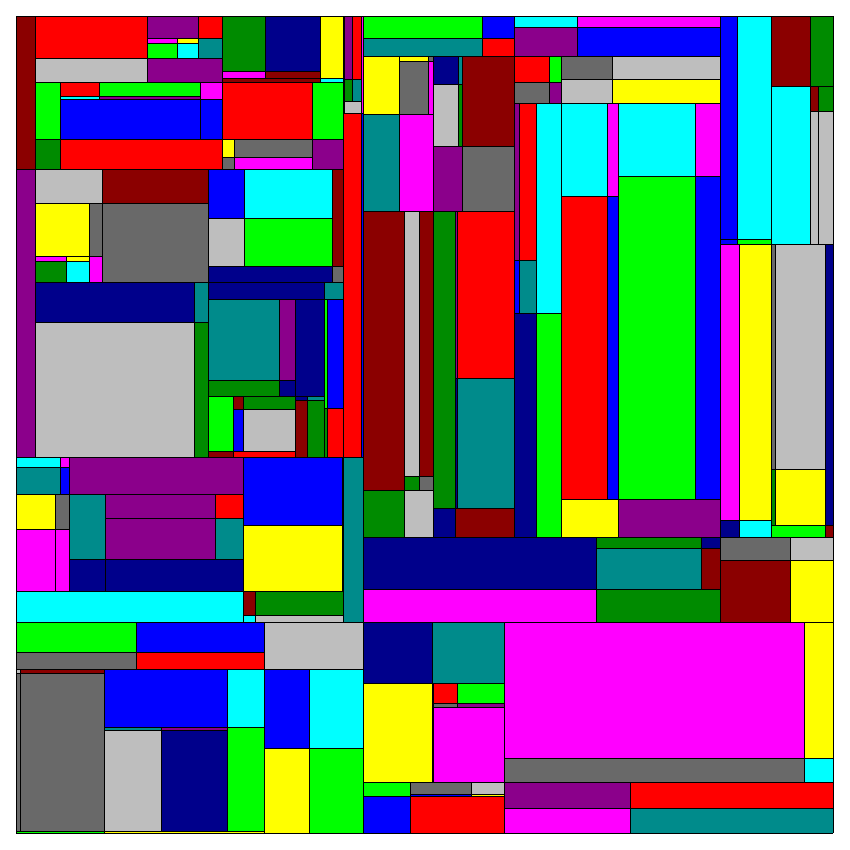}~~~
		\includegraphics[width=.2\hsize]{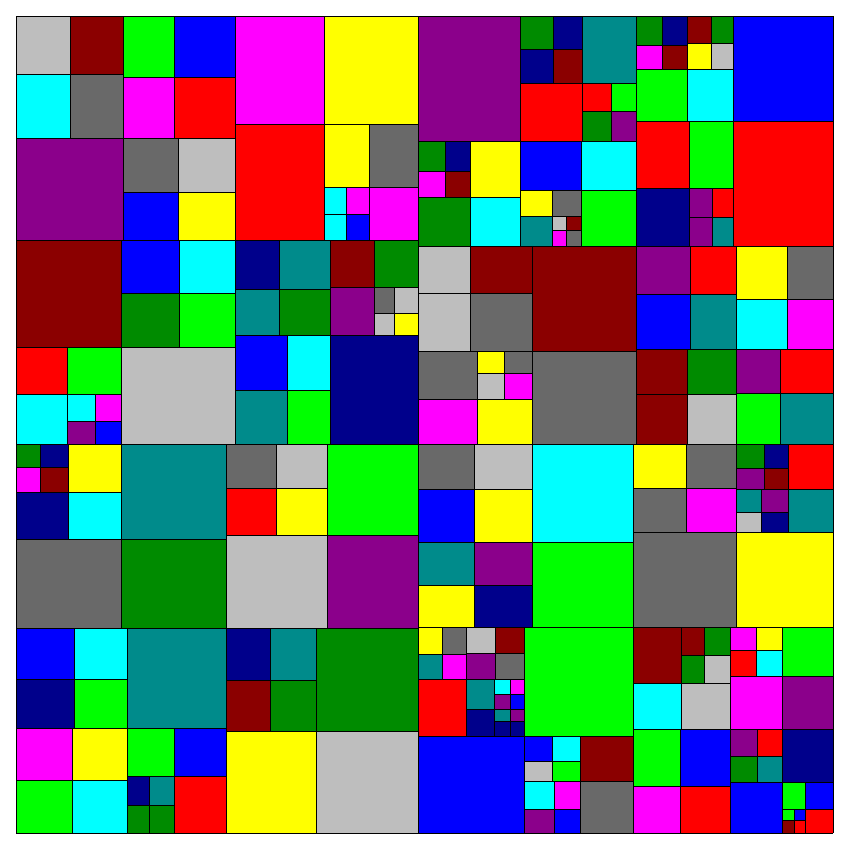}~~~
		\includegraphics[width=.2\hsize]{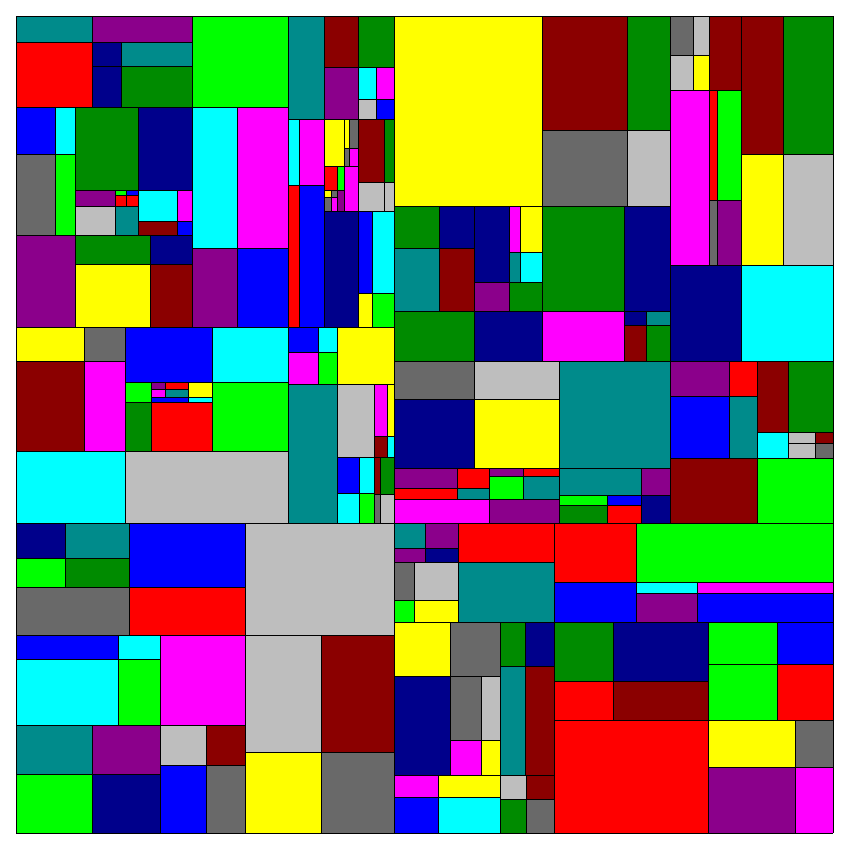}~~~
		\includegraphics[width=.2\hsize]{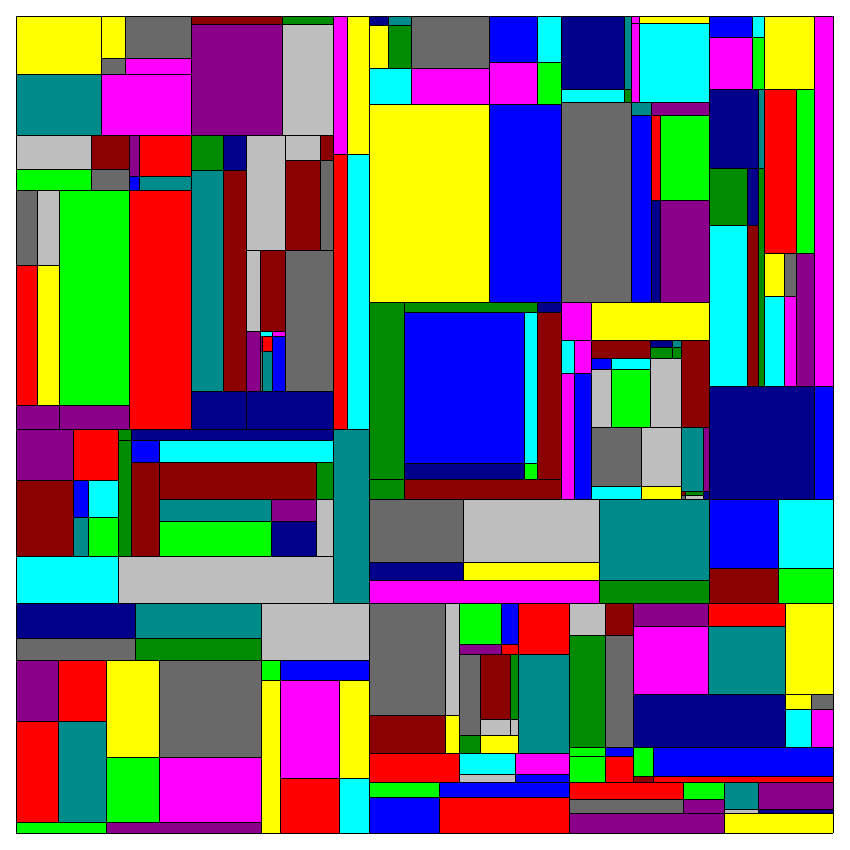}
		~\\
		\subfigure[\label{fig:wps}]{\includegraphics[width=0.2\hsize]{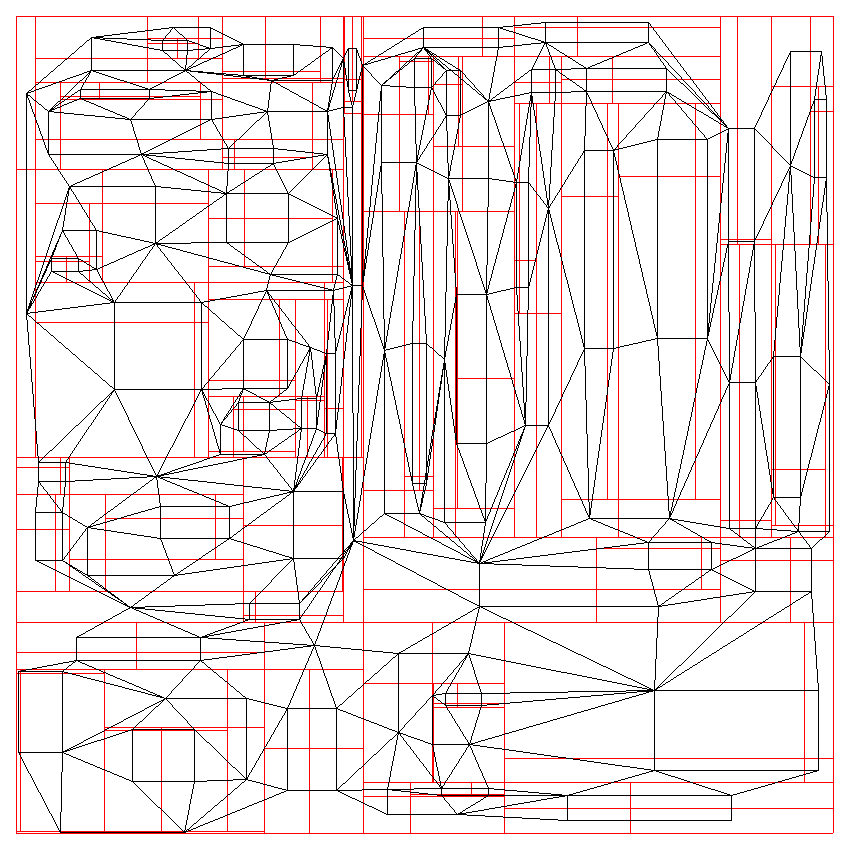}}~~~
		\subfigure[\label{fig:wpsl01}]{\includegraphics[width=.2\hsize]{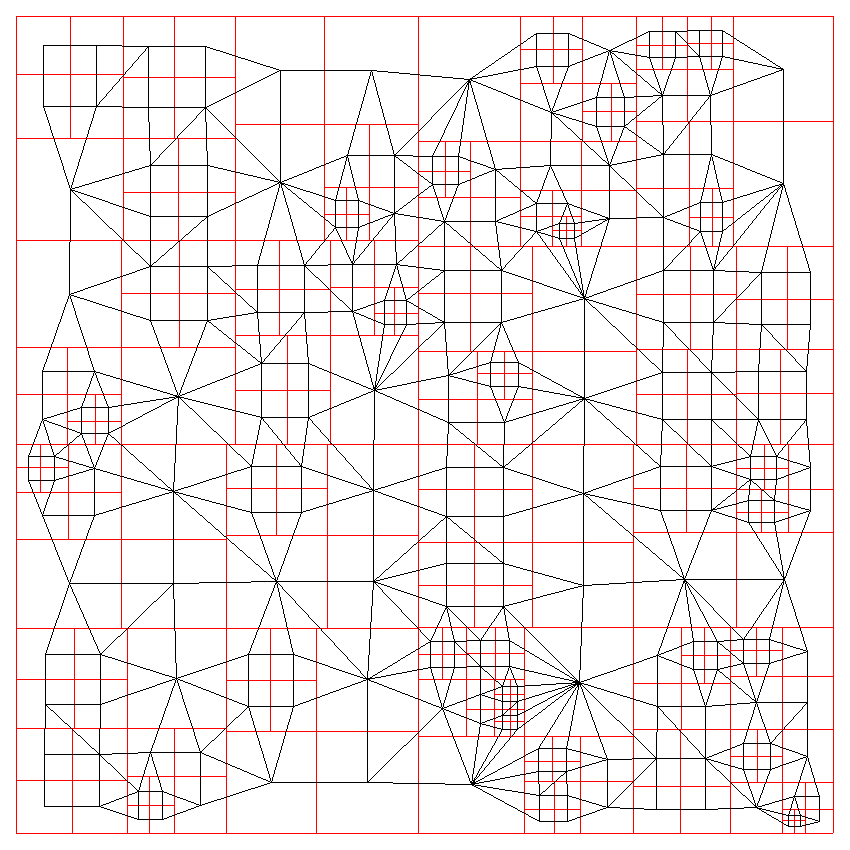}}~~~
		\subfigure[\label{fig:wpsl05}]{\includegraphics[width=.2\hsize]{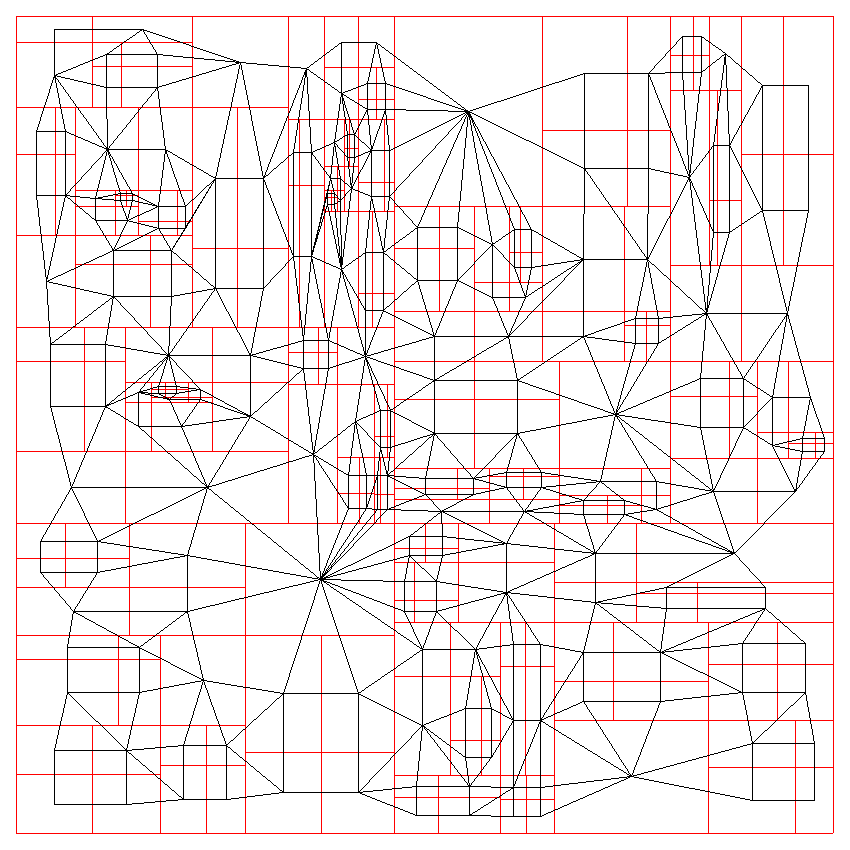}}~~~
		\subfigure[\label{fig:wpsl09}]{\includegraphics[width=.2\hsize]{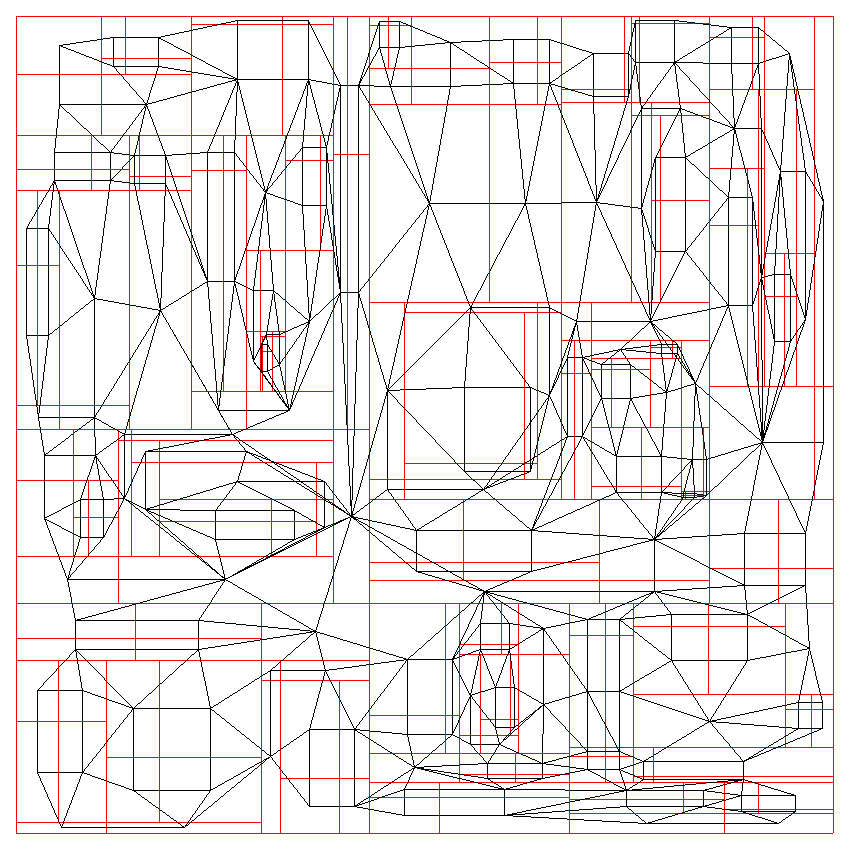}}~		
	\end{center}
	\caption{\label{fig:block}Illustration of block diagrams (top panels) and the dual network (bottom panels) obtained using the (a) original WPS ($\epsilon=1$)  and the modified one with  (b) $\epsilon = 0.1$, (c) $0.5$,  and (d) $0.9$  after $n = 85$ iterations.}
\end{figure*}

Figure~\ref{fig:wps} shows a snapshot of a random WPS lattice generated after $n=85$ iterations (256 nodes). The WPS lattice presents disorder due to its block's multifractal structure, which leads to a coordination number with a distribution that follows a power law, $P(q)\sim q^{-\alpha}$, with $\alpha=5.6$~\cite{Hassan2010}. Figures~\ref{fig:wpsl01}-\ref{fig:wpsl09} illustrate different block patterns obtained using three values of $\epsilon =0.1$, 0.5, and 0.9 after $n=85$ iterations of the rule. The bottom panels
of the figure~\ref{fig:block} show the corresponding dual networks obtained from the block diagrams of the top panels.

\begin{figure}[h]
	\subfigure[]{\includegraphics[width=.5\hsize]{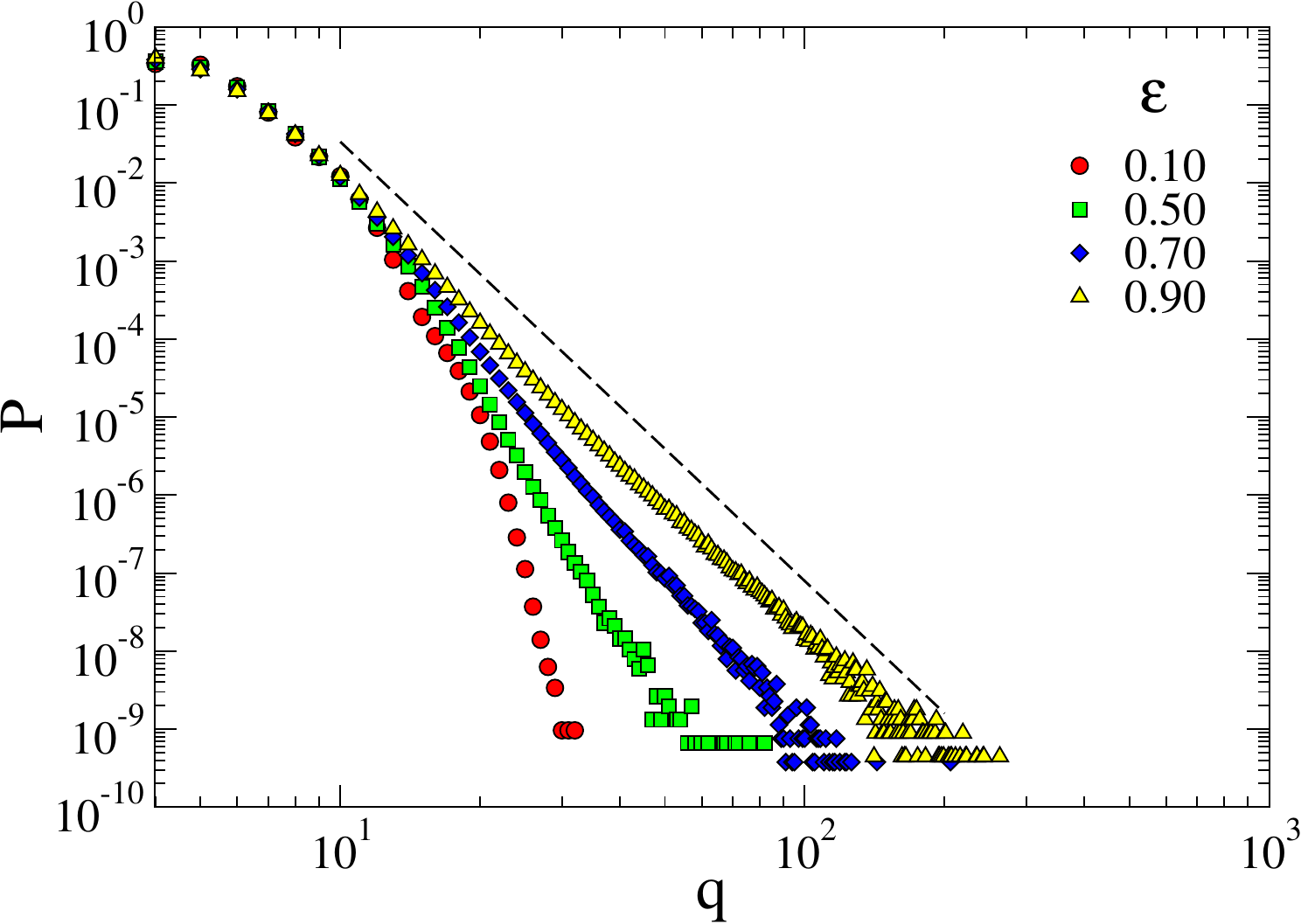}}
	\subfigure[]{\includegraphics[width=.5\hsize]{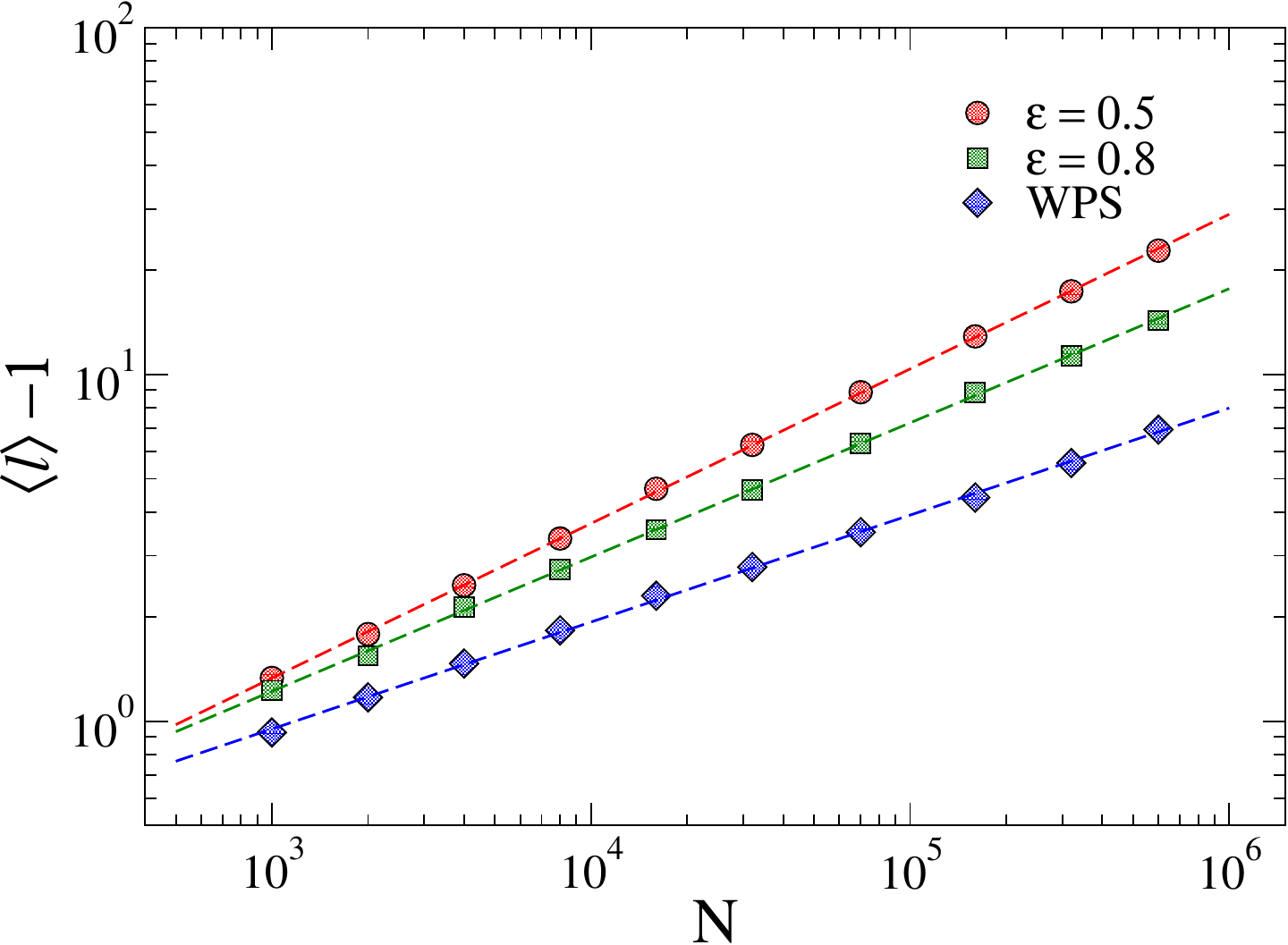}}
	\caption{\label{fig:pq} (a) Degree probability distribution for the dual WPS lattices obtained for various values of $\epsilon$. The dashed lines is a power law with exponent equals to $-5.63$. (b) Average shortest path as a function of the graph size for different values of $\epsilon$. Dashed lines are power-law regressions with slopes $1/d_t=0.443$, $0.385$  and $0.308$ for $\epsilon=0.5$, $0.8$, and $1$ (WPS), respectively.
	}
\end{figure}

The degree distributions obtained with the modified algorithm are presented in Fig.~\ref{fig:pq}(a). The networks have degree exponents varying with $\epsilon$, which interpolate between the original WPS model with exponent $\alpha\approx 5.6$ and a regular distribution of a square lattice. For $\epsilon\ll 1$, the distributions become Poissonian and lose their power-law tail. Indeed, large decay exponents $\alpha\gg 1$ are consistent with an exponential tail. We numerically calculated the topological dimension $d_\text{t}$ of the WPS model, defined in terms of the average shortest path $\langle l \rangle$~\cite{PastorSatorras2015}  as 
\begin{equation}
\langle l \rangle -1 \sim N^{1/d_\text{t}},
\end{equation}
and observed that it varies smoothly from $1/d_\text{t}=1/2$ for $\epsilon=0$ (square lattice) to $1/d_\text{t}=0.308(3)$ for $\epsilon=1$ (WPS); see Fig.~\ref{fig:pq}(b).

\begin{figure}[h]
\subfigure[]{\includegraphics[width=0.5\hsize]{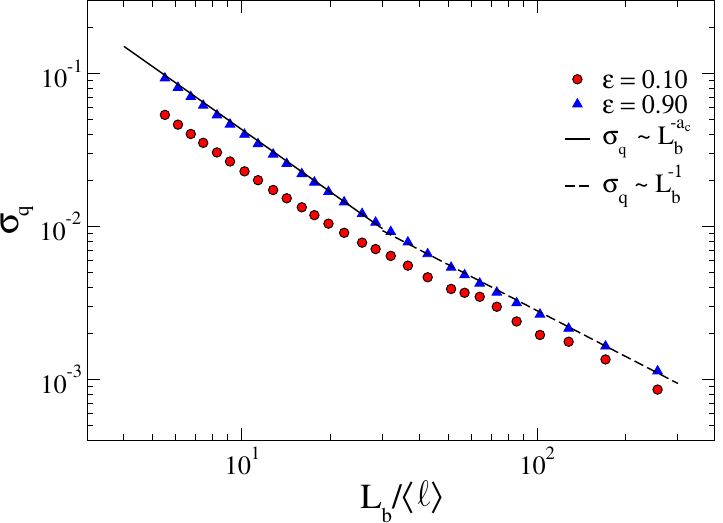}}
\subfigure[]{\includegraphics[width=0.5\hsize]{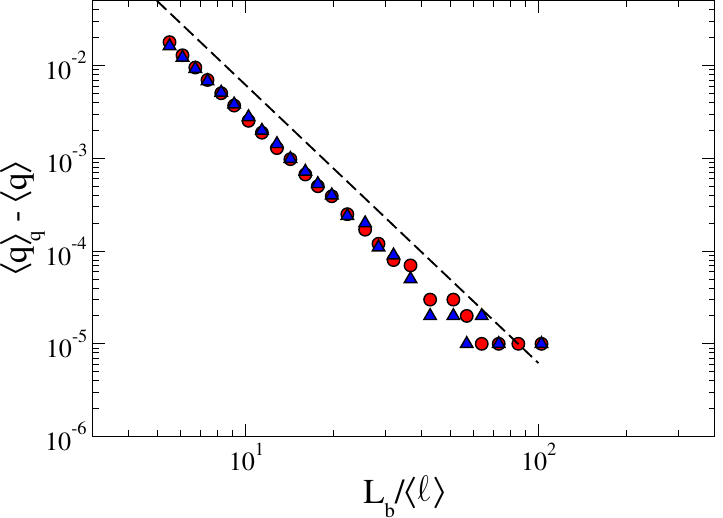}}
	\caption{\label{fig:sigma} (a) Fluctuation of the mean degree (or connectivity) as a function
		of the grid size $L_\text{b}/\langle\uell \rangle$. The solid and dashed  lines represent power-law decays with exponent $a=a_c=1.363$ and $a=1$, respectively. (b) Difference between average degree computed for blocks of size $L_\text{b}$ and the full network. Dashed line is a power-law decay $\langle q \rangle_b - \langle q \rangle\sim L_\text{b}^{-3}$.}
\end{figure}

We also investigate how the fluctuation of the average connectivity scales with the parameter $\epsilon$ using the wandering exponent obtained in the following way: The system is divided into blocks of size $L_\text{b}$, and the block-averaged standard deviation $\sigma_q$ in relation to the overall average degree $\langle q \rangle$ is calculated~\cite{Barghathi2014}. Figure~\ref{fig:sigma}(a) shows $\sigma_q$ as a function of the grid size $L_\text{b}/\langle \uell \rangle$. The exponent $a=d(1-\omega)$ of the scaling law $\sigma_q\sim L_\text{b}^{-a}$ presents an asymptotic decay for $L_b/ \langle \uell \rangle \rightarrow \infty$ consistent with uncorrelated disorder for which $a=1$. The corresponding wandering exponent satisfies the Luck-Harris criterion for models in the DP class, where disorder is a relevant perturbation of the dynamics, when $a<=a_\text{c}=1.363$. A curious behavior is present in the initial decay, where an exponent close to the critical value $a_c=1.363$ is reported. To the best of our understanding, this is a coincidence. Finally, we also analyzed how the average degree with blocks of size $L_\text{b}$, $\langle q \rangle_b$, approaches the global $\langle q \rangle$ computed for the whole lattice, presented in figure~\ref{fig:sigma}(b). A decay given by $\langle q \rangle_b - \langle q \rangle\sim L_\text{b}^{-3}$ was observed for all values of $\epsilon\in [0.1,0.9]$. Since this decay is much faster than that of the degree fluctuations, the pitfall indicated in Ref.~\cite{Barghathi2014}, where the choice of $\langle q \rangle_b$ or $\langle q \rangle$ can alter the wandering exponent, does not apply to our current analysis.

\subsection{Contact process}

The CP is a stochastic interacting particle system defined on the top of a lattice \cite{Marro1999,Harris1974}. So, each lattice node $i$ can be in one of the two states: occupied by a particle ($\sigma_i(t) = 1$) or empty ($\sigma_i(t) = 0$). The stochastic process evolves via a catalytic creation of particles: a new particle is created at a node $i$ with rate $\lambda r$, where $r$ is the fraction of occupied nearest-neighbors of $i$. The CP dynamics also include particle spontaneous annihilation with a rate $\mu=1$, independently of the neighboring sites. If all nodes are empty, the system is trapped in the absorbing state. Exactly at $\lambda=\lambda_c$, the system undergoes a continuous phase transition from the active phase to an absorbing phase within the directed percolation universality class.

The CP simulations on graphs can be implemented as follows~\cite{Marro1999} (See also \cite{Cota2017}): Randomly, select an occupied site. Then, choose between (i) annihilation, with probability $1/(1 + \lambda)$, or (ii) creation, with probability $\lambda/(1 + \lambda)$. If annihilation is selected, the node is vacated. On the other hand, if creation is chosen, one of the $q$ nearest-neighbor nodes of the selected node is chosen at random, and if empty, it becomes occupied. Otherwise, no change of state happens. The time increment associated with each event of creation or annihilation is given by $\Delta t = -\ln \xi /[(\mu+\lambda)N_\text{p}]$, where $N_\text{p}$ is the number of occupied nodes just before the attempted transition and $\xi$ is a random number uniformly distributed in the interval $(0,1)$.

\section{Results and discussion}
\label{sec:result}

We begin our study by analyzing the evolution of the number of active nodes $n(t)$. This spreading analysis is done by considering the initial condition with a single particle. The critical value of the control parameter $\lambda_\text{c}$ is defined as the smallest $\lambda$ supporting asymptotic growth of $n(t)$~\cite{Marro1999}. We found the critical value depends slightly on the parameter $\epsilon$; see Table~\ref{tab:crit}. In this part, we used lattices with $10^6$ iterations and took an average over $5 \times 10^3$ lattices, with 10 runs in each one.

\begin{figure}[h]
	\includegraphics[width=0.5\hsize]{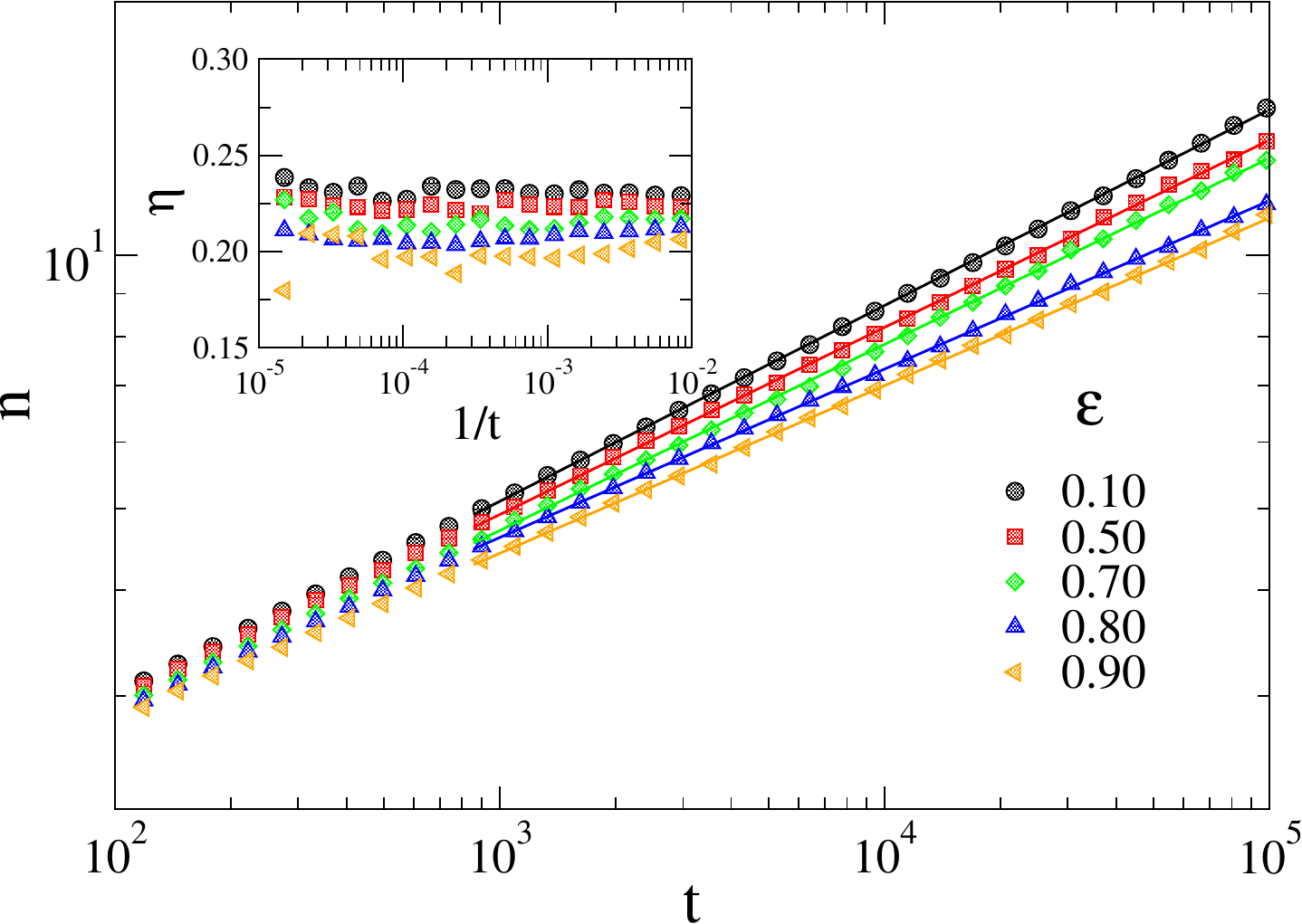} ~~~
	\caption{\label{fig:nt} Number of active sites as a function of time in the spreading analysis. We consider  different values of the control parameter. The lines are power-law regressions (Pearson regression coefficient $r>0.999$) of the data represented by symbols.  Inset shows the local slope analysis where $\eta= d\ln n / d\ln t$. The averages were computed  over 1200  network samples with $10^4$ runs each.}
\end{figure}

In Fig.~\ref{fig:nt}, we show the time evolution of the number of particles $n(t)$ at $\lambda=\lambda_\text{c}$ for different values of $\epsilon$. We observe that the asymptotic evolution follows a power law, $n(t) \propto t^\eta$, as expected for critical spreading~\cite{Marro1999,Henkel2008}. The exponent $\eta$ varies continuously with $\epsilon$, as shown in Table~\ref{tab:crit}. The critical exponents for the smallest values of $\epsilon$ are closer to the exponent $\eta= 0.2295(10)$ of the DP class~\cite{Henkel2008}. Conversely, the value $\eta= 0.185(5)$ for the original WPS lattice~\cite{Alves2022} is recovered when $\epsilon \to 1$.

\begin{figure}[h]
	\includegraphics[width=0.5\hsize]{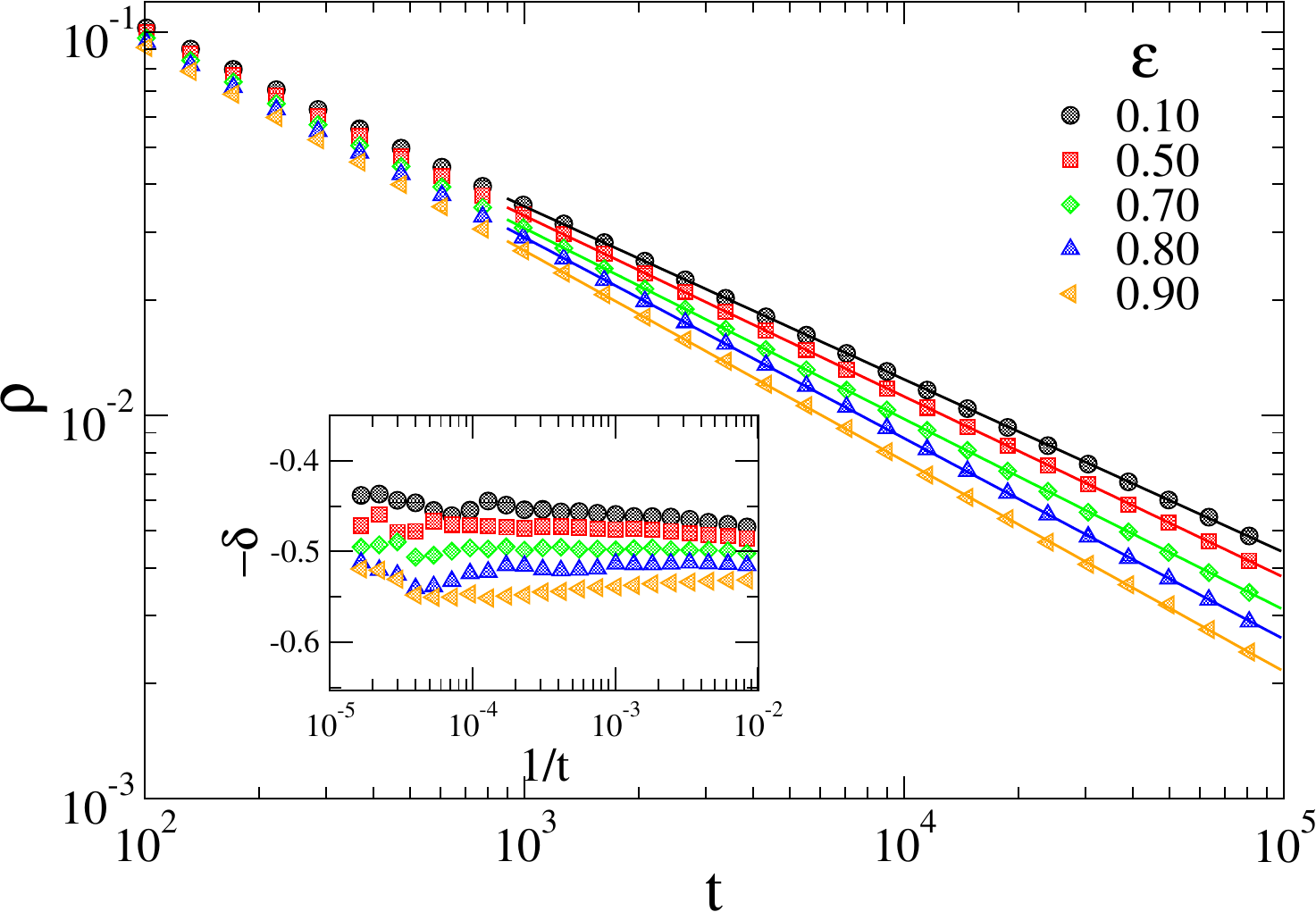} ~~~
	\caption{\label{fig:decai} Density of active sites as a function of time in the decay analysis.  The lines are power-law regressions (Pearson regression coefficient $r>0.999$) of the data represented by symbols. Inset shows the local slope analysis where $-\delta= d\ln \rho / d\ln t$. An average over $100$ lattice with 10 runs each was used.}
\end{figure}

Figure \ref{fig:decai} presents the decays of the order parameter for simulations with a fully active initial condition ($\rho(0)=1$). At criticality, one expects that the density of active sites $\rho(t)$ decays as~\cite{Marro1999} $\rho\propto t^{-\delta}$. We found continuously varying values of the critical exponents $\delta$, interpolating between the DP and original WPS exponents,  $\delta=0.4505(1)$~\cite{Henkel2008} and $\delta=0.57(3)$~\cite{Alves2022}, respectively; Table~\ref{tab:crit}. 

\begin{table}
	\begin{tabular}{c c c c c c}  \hline\hline
		& $\lambda_c$ & $\eta$ & $\delta$ & $\beta/\nu_\perp$ & $z$  \\\hline
		DP        &  - & 0.2295(10) & 0.4505(1) & 0.797(3) &  1.7674(6)  \\\hline
		$\epsilon$=0.10 &  1.5703(2)  & 0.232(5)  & 0.451(7) & 0.80(2) & 1.76(3)  \\
		0.50            &  1.5675(2)  & 0.224(5)  & 0.471(5) & 0.81(2) & 1.68(4)  \\
		0.70            &   1.5637(1) & 0.214(5)  & 0.497(4) & 0.83(2) & 1.60(3)  \\
		0.80            & 1.5611(1)   & 0.207(5)  & 0.522(8) & 0.84(2) & 1.58(3)  \\
		0.90            &  1.5577(1)  & 0.199(6)  & 0.54(1)  & 0.87(3) & 1.56(3)  \\
		WPS & 1.5525(5)  & 0.185(5) & 0.57(1) &0.84(1) & 1.59(1) \\  \hline \hline
	\end{tabular}
	\caption{Critical rates and exponents for different values of the parameter $\epsilon$. The uncertainties in last significant digit, indicated in parenthesis, were determined using a local slope analysis for the range $64\le L\le 512$ for QS and $10^3<t<10^5$ for the dynamical quantities.}
	\label{tab:crit}
\end{table}

We performed extensive simulations of the CP on the modified WPS lattices with $L=\sqrt{N}= 8, 16, 32,..., 512$, where $N=3n+1$ is the number of nodes, employing the quasi-stationary (QS) simulation method~\cite{deOliveira2005}. It is based on maintaining and gradually updating a set of configurations visited during the evolution. If a transition to the absorbing state is imminent, the system is instead placed in one of the saved configurations. This procedure has been successfully applied to investigate critical dynamics on different disordered graphs~\cite{deOliveira2008b,deOliveira2016,Schrauth2018,Ferreira2011}. Each realization of the process was initialized with all sites occupied and run for at least $10^6$ time units. Averages were taken in the QS regime after discarding an initial transient. Typically, we performed averages over at least 100 distinct lattices for each size.

\begin{figure}[h]
	\includegraphics[width=0.5\hsize]{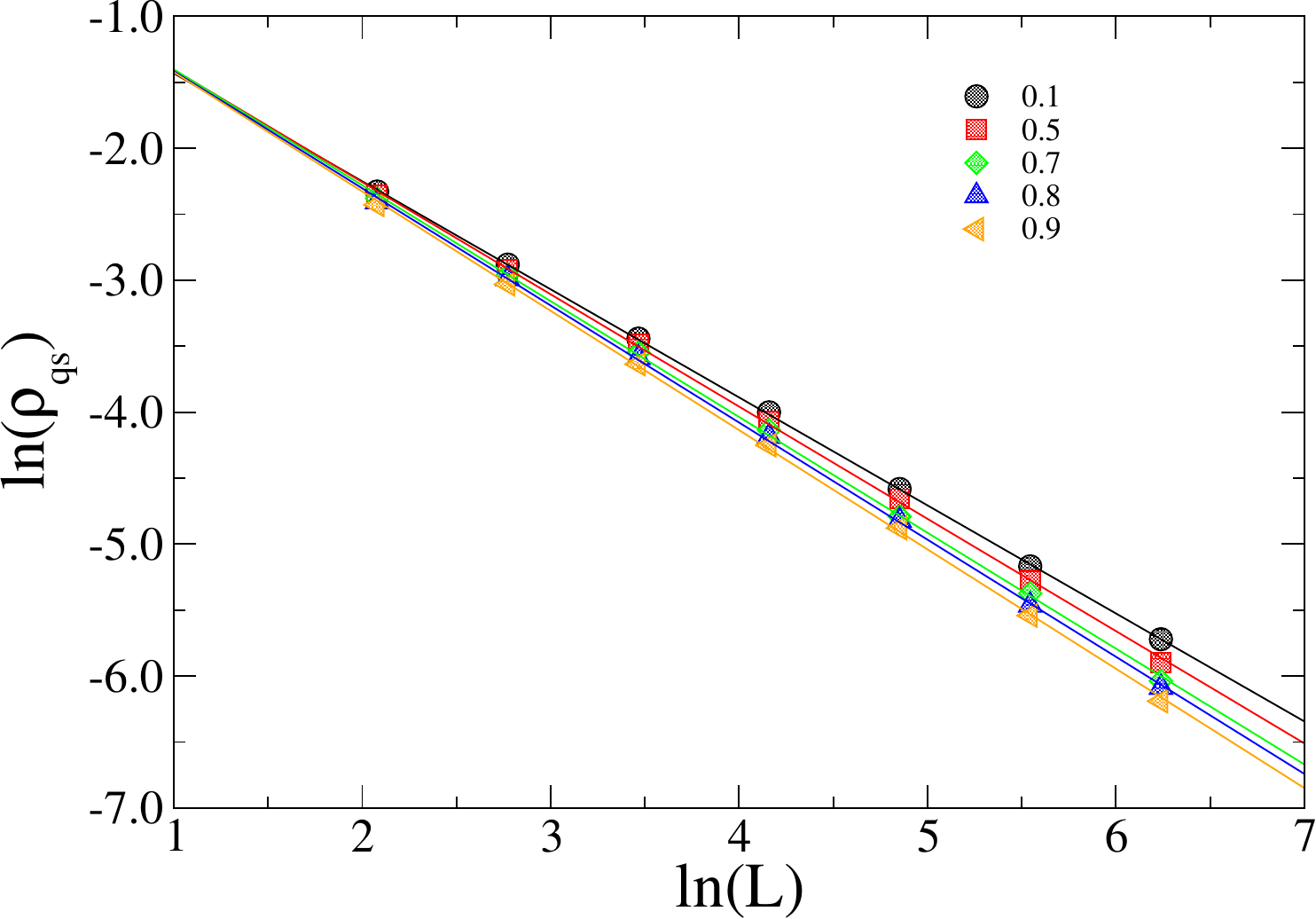} ~~~
	\caption{\label{fig:rho_qs} Double-logarithm plots of the critical QS density of active sites $\rho$ as a function of the $L=\sqrt{N}$ for different values of $\epsilon$.  Lines are linear regressions (Pearson regression coefficient $r>0.999$) of data represented by symbols.}
\end{figure}

\begin{figure}[h]
	\includegraphics[width=0.5\hsize]{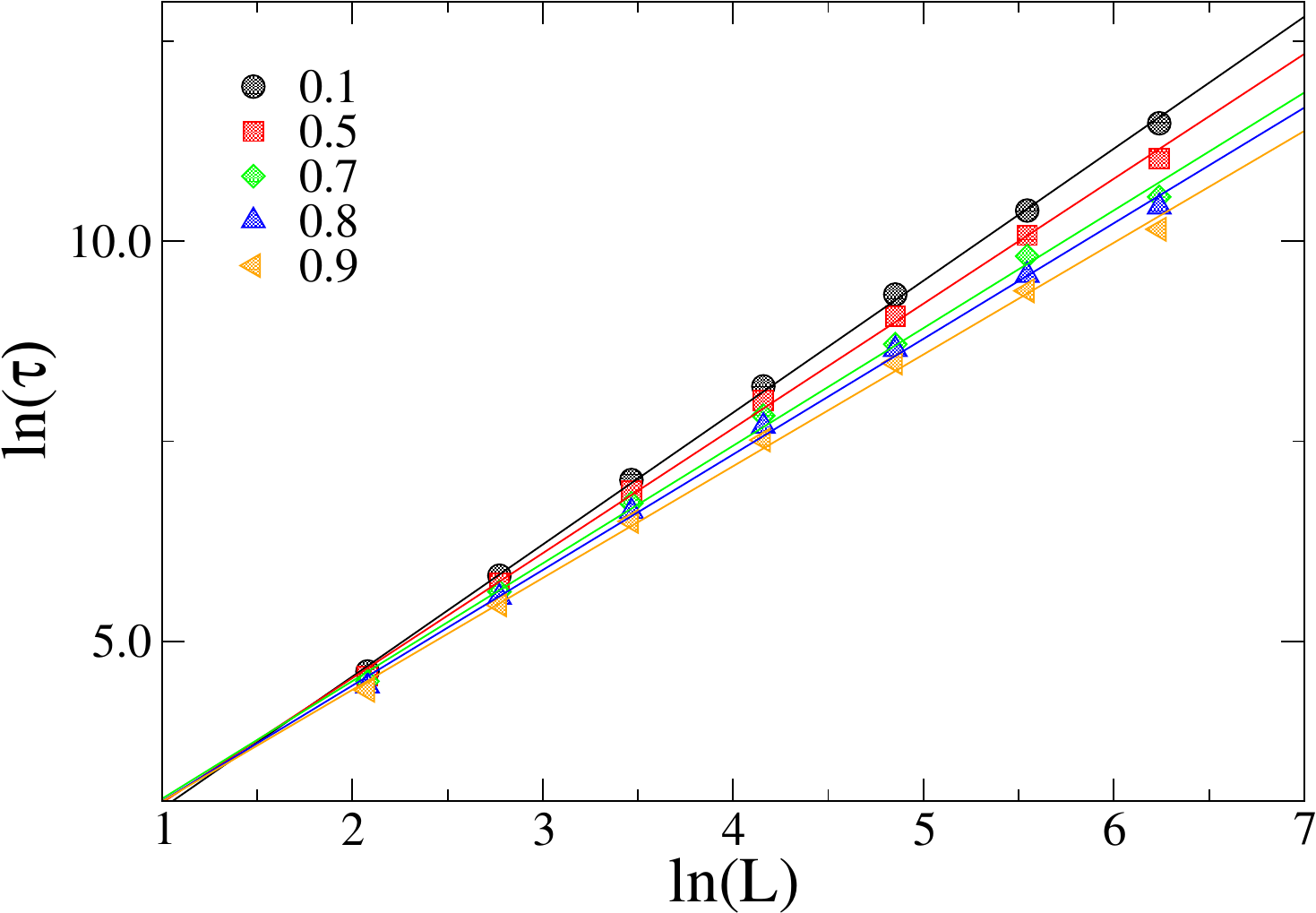} ~~~
	\caption{\label{fig:tau_n} Double-logarithm plots of the critical lifespan $\tau$ of the QS regime as a function of the $L=\sqrt{N}$ for different values of $\epsilon$. Lines are linear regressions (Pearson regression coefficient $r>0.999$) of data represented by symbols.}
\end{figure}

At $\lambda=\lambda_c$, the QS density of active sites, $\rho$, follows a power-law
$\rho\sim  L^{-\beta/\nu_\perp}$. Also, the lifetime of the QS state, $\tau$, taken as the mean time between two attempts to visit the absorbing state in the QS simulation~\cite{deOliveira2005}, follows 
$\tau\sim  L^z$ at the critical point.  The scaling laws presented in Figs.~\ref{fig:rho_qs} and \ref{fig:tau_n} lead to the exponents $\beta/\nu_\perp$ and $z$ listed in Table~\ref{tab:crit}. Analogously to the time-dependent analysis, we observe that both  exponents vary continuously with $\epsilon$, from the values obtained for the DP class to the values for the original WPS lattice.

\section{Conclusions}
\label{sec:conclu}

We modified the algorithm proposed by Hassan et. al.~\cite{Hassan2010} to construct a weighted planar stochastic lattice and  investigate the role of planar disorder. The resulting lattice presents multifractality, and its dual network degree follows a power-law degree distribution $P(q)\sim q^{-\alpha}$ with continuously varying exponents interpolating between the WPS ($\alpha=5.6$) and square lattice. We performed extensive large-scale simulations of the contact process on the modified WPS lattices and reported critical exponents depending on the parameter $\epsilon$ that controls the degree distribution of the graph. This result is qualitatively analogous to the contact processes on random, small-world, and thus infinite-dimensional graphs with power-law degree distribution~\cite{Ferreira2011,Mata2014}. However, the relationship between the critical exponent and the degree distributions of the graphs is different. In the infinite-dimensional case, a mean-field approach where the degree heterogeneity is explicitly reckoned~\cite{Ferreira2011,Mata2014} is sufficient to explain the critical exponent observed in simulations, while the CP in the planar, two-dimensional WPS networks presents the criticality at odds with the heterogeneous mean-field approach. In particular, the standard mean-field exponent, compatible with homogeneous networks, is recovered in the heterogeneous mean-field approach for $\alpha>3$, whereas the simulation of CP on WPS lattices, which are featured by $\alpha>5.6$, is far from mean-field.

Finally, the analysis of degree fluctuations in the WPS lattices across different scales leads to an asymptotic wandering exponent $\omega=1/2$ that indicates the disorder should be a relevant perturbation according to the Luck-Harris criterion~\cite{Luck1993}. Our work reveals agreement with the criterion, with weaker effects than usually observed for the DP class under uncorrelated disorder that presents the same wandering exponent for all strengths of the disorder controlled by the parameter $\epsilon\in[0.1,0.9]$.

The present study is based on numerical simulations where scaling-laws are influenced by finite-size effects such that other phenomena could eventually be observable in the infinite-size limit. So, more analytical approaches, such as a numerical renormalization group~\cite{Vojta2009}, could shed light on these questions. Prospectively, investigating the structural properties that govern the relevance of the disorder should be addressed in a future work.


\section{acknowledgement}
SGA thanks the support of CNPq (Grant no.  	311019/2021-8). MMO thanks the support of CNPq (Grant no. 312462/2021-2) and  FAPEMIG (Grant no. APQ02393-18). SCF thanks the support of CNPq (Grant no. 310984/2023-8).

%

\end{document}